\pdfoutput=1
\documentclass[11pt,a4paper]{article}
\usepackage[hmargin=25mm,vmargin=25mm]{geometry}
\usepackage[T1]{fontenc}
\usepackage{newtxtext,newtxmath}
\usepackage{graphicx}
\usepackage{amsmath}
\usepackage{array}
\usepackage{booktabs}
\usepackage{xcolor}
\usepackage{placeins}
\usepackage{siunitx}
\usepackage{url}
\usepackage[labelfont=bf,labelsep=period,font=small]{caption}
\usepackage{cite}
\usepackage{hyperref}
\input{si-xr}
\hypersetup{colorlinks=true,
  linkcolor={blue!55!black}, citecolor={blue!55!black},
  urlcolor={blue!55!black}, filecolor={blue!55!black},
  breaklinks=true, bookmarksnumbered=true, pdfstartview={FitH},
  pdftitle={Autonomous discovery of new structure-plausibility laws for explainable and rapid crystal diagnosis and screening},
  pdfauthor={Zhilong Song, Heung Law, Lixue Cheng}}

\newcommand{\rhoc}{\ensuremath{\rho}}
\newcommand{\fion}{\ensuremath{f_{\mathrm{i}}}}
\newcommand{\VM}{\ensuremath{E_{\mathrm{M}}}}

\makeatletter
\renewcommand{\abstract}[2][]{\gdef\PRIS@abstract{#2}}
\newcommand{\keywords}[1]{\gdef\PRIS@keywords{#1}}
\newcommand{\PRISabstract}{\PRIS@abstract}
\newcommand{\PRISkeywords}{\PRIS@keywords}
\makeatother

\begin{document}

\title{Autonomous discovery of new
structure-plausibility laws for explainable and rapid crystal diagnosis and screening}

\abstract[ABSTRACT]{Generative models and autonomous agents now propose candidate inorganic crystals by the million, yet most screens test little beyond a minimum-distance cutoff and give no chemical reason for failure.
For nearly a century, Pauling's rules have served as the main empirical theory of crystal-structure stability, but they remain qualitative and idealized, leading to only 6.5\% of known ionic crystals satisfying rules 2--5 simultaneously.
Here our agents generate, test and refute two million candidate laws to discover eight Plausibility Rules for Inorganic Structures (PRIS).
These laws encode five physical mechanisms: short-range repulsion, ionic contact and packing, electrostatic balance, bond-valence conservation and crystallographic site simplicity.
Experimental structures satisfy the PRIS law sets in 82--99\% of cases, and the strictest set detects 87.9\% of artificially damaged structures, far above the 1.6--3.2\% of conventional distance cutoffs.
Although derived without synthesis labels, PRIS plausibility is highly linearly correlated with synthesizability, and a continuous PRIS-derived synthesis score (PSS) explainably filters out 83.7\% of hard-to-synthesize structures while retaining 80.7\% of experimental structures.
In an inverse-design task targeting ultra-incompressible crystals, explainable screening by PRIS and PSS cuts the density functional theory (DFT) validation queue by up to 67.3\% while retaining 99.2\% of the DFT-verified target candidates.
Beyond screening, PRIS traces GNoME's low-symmetry excess to artificial element ordering and identifies falsified crystal structures.
PRIS moves screening from a pass-or-fail verdict to a chemical reason for failure, showing that autonomous agents can discover, by active refutation, physicochemical laws that guide calculations and experiments.}


\makeatletter
\begin{center}
{\LARGE\bfseries \@title\par}
\bigskip
{\large Zhilong Song$^{1,2}$, Heung Law$^{1,2}$ and Lixue Cheng$^{1,2,3,*}$\par}
\medskip
{\small
$^{1}$Department of Chemistry, Hong Kong University of Science
and Technology, Kowloon, Hong Kong 999077, China\par
$^{2}$IAS Center for AI for Scientific Discoveries, Hong Kong
University of Science and Technology, Kowloon, Hong Kong 999077, China\par
$^{3}$Department of Chemical and Biological Engineering, Hong
Kong University of Science and Technology, Kowloon, Hong Kong 999077, China\par
$^{*}$Corresponding author: \href{mailto:lixuecheng@ust.hk}{lixuecheng@ust.hk}\par}
\end{center}
\makeatother

\begin{quote}
\small\noindent\textbf{Abstract.} Generative models and autonomous agents now propose candidate inorganic crystals by the million, yet most screens test little beyond a minimum-distance cutoff and give no chemical reason for failure.
For nearly a century, Pauling's rules have served as the main empirical theory of crystal-structure stability, but they remain qualitative and idealized, leading to only 6.5\% of known ionic crystals satisfying rules 2--5 simultaneously.
Here our agents generate, test and refute two million candidate laws to discover eight Plausibility Rules for Inorganic Structures (PRIS).
These laws encode five physical mechanisms: short-range repulsion, ionic contact and packing, electrostatic balance, bond-valence conservation and crystallographic site simplicity.
Experimental structures satisfy the PRIS law sets in 82--99\% of cases, and the strictest set detects 87.9\% of artificially damaged structures, far above the 1.6--3.2\% of conventional distance cutoffs.
Although derived without synthesis labels, PRIS plausibility is highly linearly correlated with synthesizability, and a continuous PRIS-derived synthesis score (PSS) explainably filters out 83.7\% of hard-to-synthesize structures while retaining 80.7\% of experimental structures.
In an inverse-design task targeting ultra-incompressible crystals, explainable screening by PRIS and PSS cuts the density functional theory (DFT) validation queue by up to 67.3\% while retaining 99.2\% of the DFT-verified target candidates.
Beyond screening, PRIS traces GNoME's low-symmetry excess to artificial element ordering and identifies falsified crystal structures.
PRIS moves screening from a pass-or-fail verdict to a chemical reason for failure, showing that autonomous agents can discover, by active refutation, physicochemical laws that guide calculations and experiments.
\end{quote}
\medskip

\section{Introduction}

Materials design now produces candidate crystals far faster than they can be validated.
High-throughput databases, crystal generators and tool-using agents
\cite{jain2013commentary,merchant2023scaling,xie2022crystal,jiao2023crystal,sriram2024flowllm,antunes2024crystal,boiko2023autonomous,mbran2024augmenting}
supply candidates by the hundred thousand, whereas density functional
theory (DFT) energy and phonon calculations, let alone synthesis experiments, can assess only a fraction
\cite{petretto2018highthroughput,zhu2024highthroughput,szymanski2023autonomous}.
The central bottleneck has shifted from generating candidate structures to deciding which
of them are chemically reasonable crystals worth computing and synthesizing.

Making that decision requires defining what constitutes a reasonable crystal. Crystallography defines a crystal as any solid with an essentially
discrete diffraction pattern \cite{iucr1992crystal}, and every periodic cell produced by a
generative model meets that definition by construction. Chemistry demands more, and its classic
answer is Pauling's five rules \cite{pauling1929principles}. In these rules, each cation--anion contact matches the sum of ionic radii within
a coordination polyhedron set by the radius ratio, the electrostatic bond
strengths reaching each anion balance its formal charge, polyhedra avoid sharing the edges and
faces that bring cations close together, and a crystal contains few distinct kinds of
site. Alongside Goldschmidt's radius rules and structural-hierarchy principles
\cite{goldschmidt1926gesetze,pettifor1984chemical,zunger1980systematization,hawthorne2014structure},
Pauling's rules remain the main empirical theory of crystal-structure stability
\cite{george2020limited}. Their descendants are in daily use: bond valences validate
crystal refinements and map ion-migration paths in solid electrolytes
\cite{brown2009recent,wong2021softbv}, while radius-based tolerance factors screen
perovskites \cite{bartel2019new}.

Yet when applied to individual crystal structures, these
classical rules remain qualitative and idealized. In an audit of about 5{,}000 oxides, the
radius-ratio rule matched 66\% of cation environments and the electrostatic valence rule held
closely mainly in undistorted high-symmetry crystals \cite{george2020limited}. Only 13\% of those oxides
satisfied rules 2--5 together \cite{george2020limited}, and only 6.5\% of the
charge-balanced ionic experimental structures in our benchmark satisfied all four. George and
co-workers therefore called for new empirical rules beyond the nearly century-old set
\cite{george2020limited}, while crystal chemists are re-examining those principles on
quantum-mechanical and bond-topological grounds
\cite{gibbs2022pauling,hawthorne2026crystal}.

Current generative pipelines test crystal validity with far weaker filters. A proposed
structure is usually accepted whenever no two atoms lie closer than 0.5 or 0.7\,\AA{} and
its composition can be charge-balanced
\cite{court20203d,xie2022crystal,jiao2023crystal,miller2024flowmm,betala2025lematgenbench}.
In our evaluations on experimental crystals deliberately corrupted by geometric distortions
or ion swaps (hereafter artificially damaged structures), those minimum-distance cutoffs
detected just 1.6\% and 3.2\% of the damaged structures. Three recent controversies illustrate what such permissive filters
let through. GNoME reported 381{,}000 newly discovered structures on the convex hull,
expanding its stable set to 421{,}000 crystals \cite{merchant2023scaling}, yet this set remains
enriched in rare low-symmetry structures that pass distance cutoffs. Subsequent critiques traced
part of this excess to the artificial ordering of chemically similar elements, such as rare-earth pairs
and Zr--Hf, across crystallographic sites that should be equivalent
\cite{cheetham2024artificial,smit2026dataonly}. Chemical ordering also shaped A-Lab's
success criterion, which accepted ordered and partially disordered products
\cite{szymanski2023autonomous}. A reanalysis disputed several novelty and phase assignments
\cite{leeman2024challenges}, and an author correction later found four of A-Lab's 40
successes inconclusive \cite{szymanski2026correction}. An even more severe error assigns the
wrong element to valid crystallographic coordinates. Harrison and colleagues documented at least 70 falsified
crystal structures in which genuine diffraction data were paired with altered element identities
\cite{harrison2010falsified}, and an archive of retracted depositions records the same
substitution pattern across Cu, Ni, Mn and Fe labels \cite{iucr2012retraction}. Minimum-distance cutoffs
ignore both chemical ordering and elemental identity, while routine energy relaxations do not
test whether an ordered cell or elemental assignment is chemically credible. Meanwhile,
black-box synthesizability predictors based on crystal-likeness scores \cite{jang2020structure},
language models \cite{song2025llm} or recommendation engines
\cite{griesemer2025wideranging} state no explicit chemical law and give no reason for rejection. What is
missing between Pauling's idealized rules and element-blind distance cutoffs is a rapid, explainable set of
falsifiable crystal-chemical laws: quantitative bounds that experimental crystals obey, unphysical
structures violate and chemists can use to pinpoint which physical constraint has failed \cite{smit2026dataonly}.

Here we show that autonomous agents driven by active refutation can discover these laws
directly from crystallographic data. Starting from experimental structures
\cite{zagorac2019recent,grazulis2009crystallography}, interpretable crystal-chemical
quantities and prescribed ways to damage a crystal, our agents
generated, tested and actively refuted two million candidate laws.
Eight laws survived refutation to form the
Plausibility Rules for Inorganic Structures (PRIS), answering the call for new empirical
rules beyond Pauling's \cite{george2020limited}. Each law encodes one of five mechanisms: short-range repulsion, ionic
contact and packing, electrostatic balance, bond-valence conservation and crystallographic
site simplicity, so every violation identifies the mechanism that fails. Sets of these laws are satisfied by
82--99\% of experimental structures, while the full set detects 87.9\% of artificially
damaged structures. Although discovered without synthesis labels, PRIS
plausibility is linearly correlated with predicted synthesizability, and a
PRIS-derived synthesis score (PSS) explainably rejects 83.7\% of hard-to-synthesize structures while
retaining 80.7\% of experimental crystals. In an inverse-design task targeting
ultra-incompressible crystals, explainable screening by PRIS and PSS cuts the DFT validation queue by up to 67.3\% while keeping 99.2\% of the candidates that reach the design target under DFT.
The same mechanisms trace GNoME's low-symmetry excess to
artificial cation ordering and expose wrong-element assignments hidden behind plausible coordinates in falsified crystal reports.
PRIS thereby turns crystal screening from a pass-or-fail check into a chemical diagnosis, showing that autonomous agents can discover physicochemical laws that guide subsequent
calculations and experiments.

\section{Results}

\subsection{Autonomous discovery of eight laws by proposal, test and refutation}
\label{sec:loop}

\begin{figure*}[tp]
  \centering
  \includegraphics[width=0.99\textwidth]{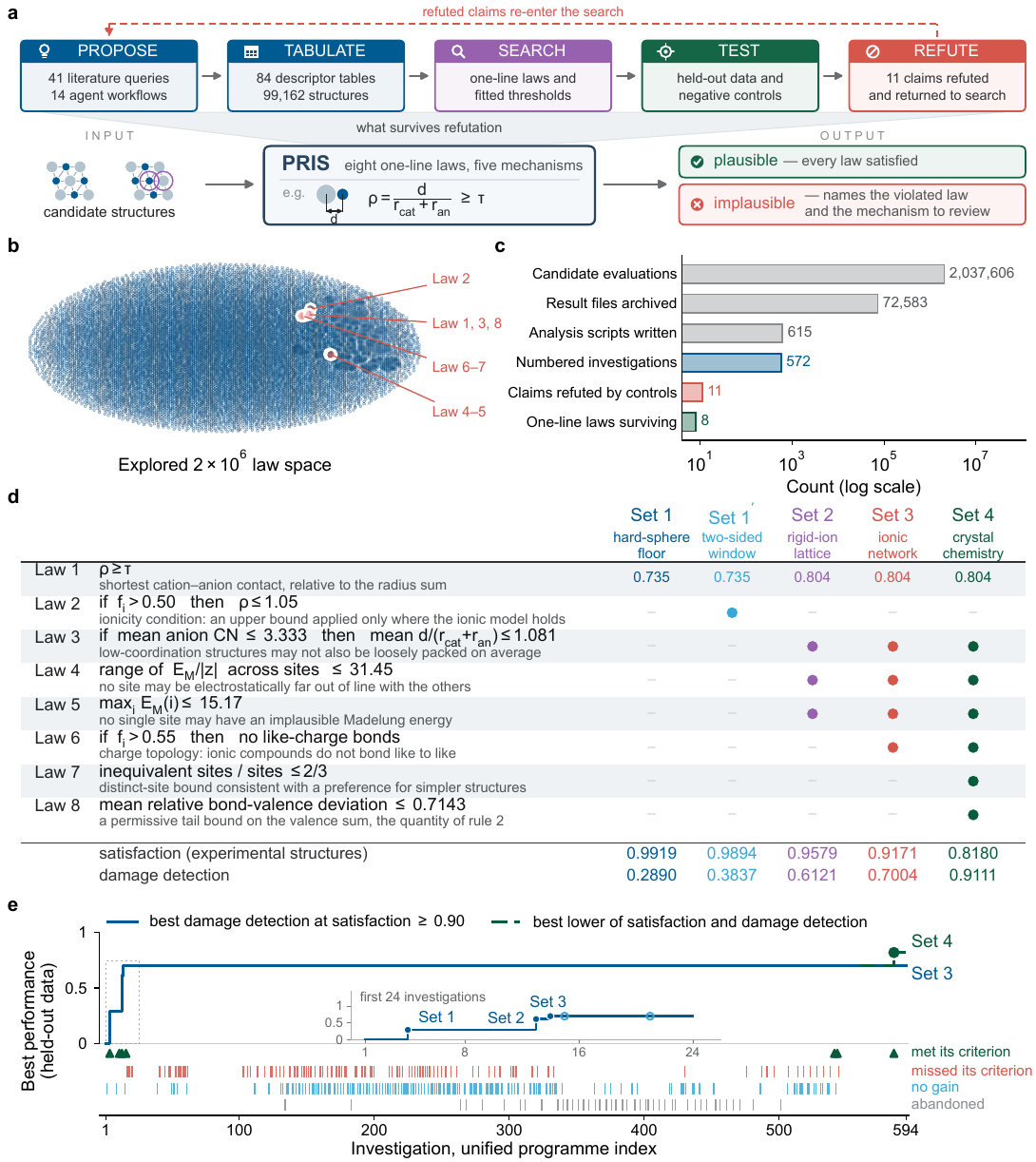}
  \vspace{16pt}
  \caption{\textbf{Autonomous discovery by proposal, testing and refutation.}
  \textbf{a},~Five-stage workflow: candidate-law proposal, descriptor tabulation,
  systematic search, held-out testing and attempted refutation. The dashed
  arrow returns refuted claims to the next cycle. Below, the eight surviving laws judge a
  candidate structure plausible or implausible. An implausible verdict
  names the violated law and its physical failure mode. Purple rings on the input
  structures mark an exchanged pair.
  \textbf{b},~t-distributed stochastic neighbour embedding (t-SNE) projection of the
  archived one-line candidate-law statements. Blue intensity encodes the number of
  statements per bin on a logarithmic scale. Red circles locate the statements nearest
  Laws~1--8.
  \textbf{c},~Counts of candidate evaluations, archived result files, analysis scripts,
  investigations, refuted claims and surviving laws. Values are on a logarithmic axis.
  \textbf{d},~Laws~1--8, their mathematical predicates (where \rhoc{} is the reduced contact of equation~\ref{eq:rho}, \(\tau\) is its threshold, \fion{} is Pauling's ionicity, CN is coordination number and \(\VM\) is the site Madelung energy) and their five nested sets.
  Each set is named after the crystal model it enforces and is shown alongside its held-out experimental satisfaction and damage detection.
  \textbf{e},~Running best held-out performance versus investigation index. Curves trace damage
  detection above the satisfaction floor (blue) and the smaller of satisfaction and
  detection (green). The inset enlarges the earliest investigations, and the strip beneath
  shows investigation outcomes.}
  \label{fig:loop}
\end{figure*}

Crystal chemistry constrains a structure through a small number of measurable quantities. A useful
crystal-chemical law must express one such quantity as a concise inequality that a chemist
can read, calculate and test, and that experimental structures obey while damaged variants
violate. Although autonomous
agents can generate candidate inequalities rapidly, turning hypothesis generation into
scientific discovery requires systematic falsification. Our five-stage workflow (Fig.~\ref{fig:loop}a) therefore separates law proposal from refutation. First, agents queried the crystal-chemical literature to propose candidate descriptors and tabulated them across experimental and damaged structures. Next, they searched for one-line inequalities and calibrated each threshold on a training subset (the discovery split). Finally, they tested surviving candidates on a separate test subset (the held-out split) and
designed adversarial counterexamples and physical controls to refute their own claims.
Every refuted claim remained in the archive to guide subsequent search cycles,
turning the agent workflow into a cumulative sequence of falsifiable experiments (Supplementary
Figs.~S1 and~S2).

We drew fixed data splits from 99{,}162 experimentally determined crystal structures in the
Inorganic Crystal Structure Database (ICSD) and the Crystallography Open Database (COD) \cite{zagorac2019recent,grazulis2009crystallography} (details in Methods and Supplementary
Fig.~S3). To construct the experimental ionic set, we selected compounds of at least two elements whose compositions can be charge-balanced with integer formal oxidation states, designating positively and negatively charged species as cations and anions. Each artificially damaged structure was generated
from an experimental parent crystal by applying one of five perturbations: 15--30\% uniaxial compression, swapping cations of different formal charges, random Gaussian atomic displacement with a standard deviation of 0.3--0.8\,\AA{}, 20--40\% isotropic expansion or swapping cations with anions. Under plane-wave DFT relaxation, the five damage classes released medians of 0.30--3.15\,eV per atom, compared with 0.0009\,eV per atom for unmodified parents (Supplementary Fig.~S24b). The median energy released in every perturbation class therefore exceeded the 90th percentile of known inorganic crystalline metastability (0.067\,eV per atom \cite{sun2016thermodynamic}) by more than fourfold. We evaluated candidate laws on experimental-structure satisfaction (the fraction of experimental structures satisfying the law set) and damage detection (the fraction of damaged structures rejected by at least one law; details in Methods).

Across 572 investigations spanning plausibility, stability and synthesis, our agents
logged 2{,}037{,}606 candidate evaluations across the statement space mapped in Fig.~\ref{fig:loop}b and summarized by funnel stage in Fig.~\ref{fig:loop}c. Active refutation eliminated eleven initially promising conclusions
that stemmed from statistical class-imbalance artefacts, rules that exploited how the perturbations were generated rather than real physics, or subtle evaluation artefacts (Supplementary Section~S6).
Because every candidate had to withstand held-out testing, physical checks and adversarial controls, only four candidate law sets across all 572 investigations surpassed the held-out frontier, the
last after more than five hundred investigations (Fig.~\ref{fig:loop}e). Just eight one-line laws survived all two
million evaluations to form PRIS.
Each surviving law states one measurable quantity, the direction in which it fails and, where needed, its chemical domain.
In Fig.~\ref{fig:loop}d, the eight laws form five nested sets that demand progressively fuller models of an ionic crystal. Set~1 enforces a hard-sphere contact floor, Set~1$'$ bounds the two-sided ionic contact
window, and Set~2 models a rigid-ion lattice with packing and Madelung laws. Expanding to Set~3 treats that lattice as a charge-consistent bond network, and Set~4 adds
valence and site laws (Supplementary Section~S18).

\subsection{Balancing experimental-structure satisfaction with damage detection}
\label{sec:rules}\label{sec:core}\label{sec:d1}\label{sec:d2}\label{sec:l4}

\begin{figure*}[tp]
  \centering
  \includegraphics[width=\textwidth]{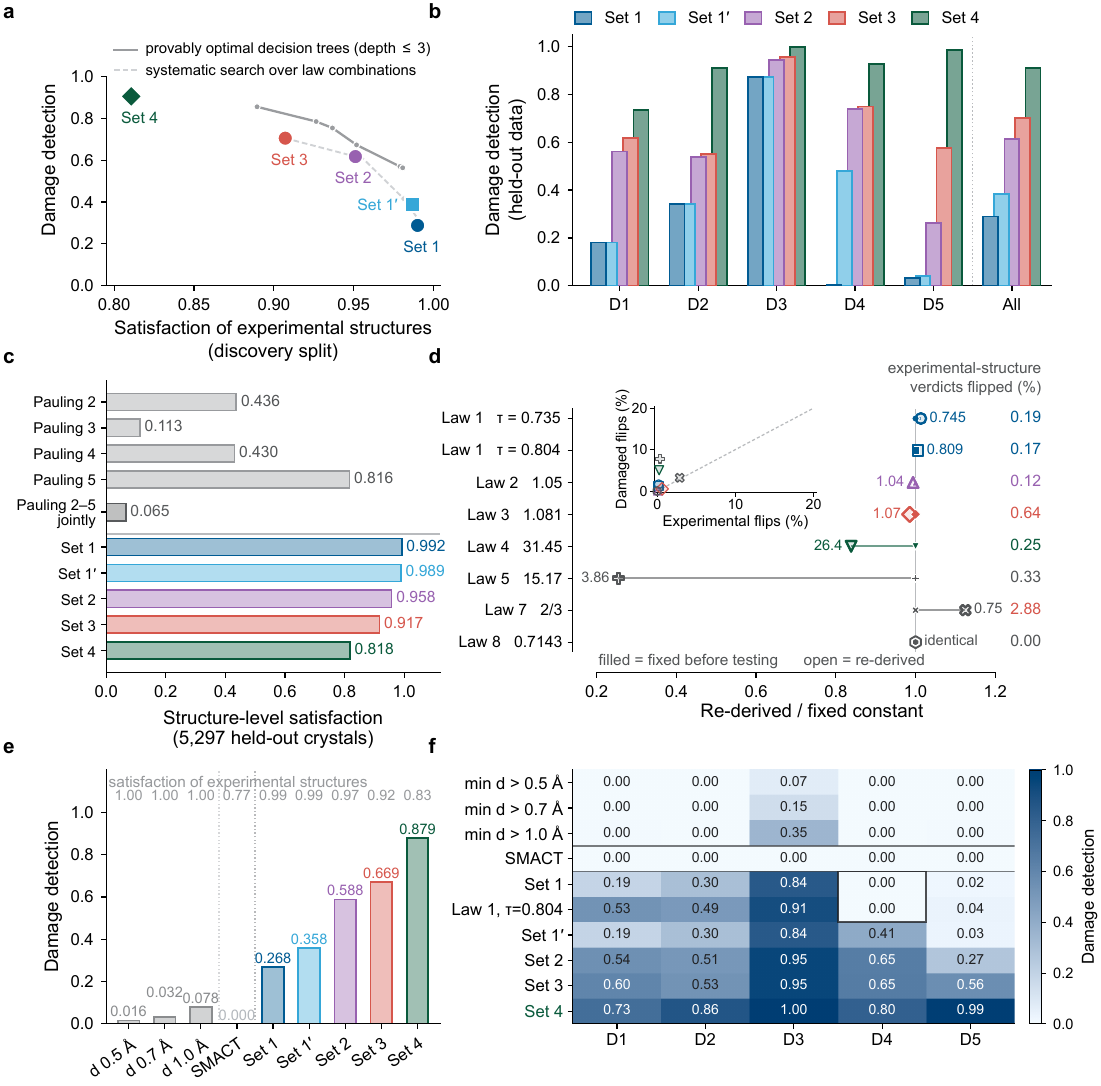}
  \caption{\textbf{Experimental-structure satisfaction, damage detection, threshold
  transfer and comparison with distance cutoffs.}
  \textbf{a},~Damage detection versus satisfaction on the discovery split. Curves plot the
  depth-limited decision-tree frontier (solid grey), interpretable law combinations (dashed grey) and the five law sets (colours).
  \textbf{b},~Held-out damage detection by composition-preserving perturbation class.
  The rightmost bar gives pooled detection (dotted divider).
  \textbf{c},~Satisfaction of Pauling rules 2--5, individually and jointly (grey), on held-out structures.
  Coloured bars give the five law sets on the same structures.
  \textbf{d},~Law thresholds re-derived at their defining percentiles on held-out data.
  Values are shown as the ratio to the frozen thresholds, with unity meaning no change. The right-hand
  column gives the percentage of held-out experimental-structure verdicts that change.
  The inset compares verdict changes for experimental and artificially damaged structures on
  a common percentage scale, where the dashed line marks equality. Marker shape identifies
  the law in both the main panel and inset.
  \textbf{e},~Overall damage detection on a common benchmark of 440 experimental structures
  and 2{,}024 damaged variants for fixed-distance cutoffs, the composition-only
  charge-neutrality check (SMACT) and successive PRIS criteria. Upper labels give
  experimental-structure satisfaction.
  \textbf{f},~Damage detection on the same benchmark by perturbation class for baselines,
  contact laws and law sets. The Law~1 row applies the \(\tau=0.804\) floor alone.
  Outlines mark where a contact floor alone misses isotropic expansion.}
  \label{fig:rules}
\end{figure*}

Any plausibility screen faces a fundamental trade-off between retaining valid experimental
crystals and rejecting damaged ones, requiring experimental-structure satisfaction and damage
detection to be evaluated jointly. On the discovery split (Fig.~\ref{fig:rules}a), the nested Sets~1--4 follow the frontier of systematic law combinations and approach that of provably optimal depth-limited decision trees. By contrast, composition-only charge neutrality detected no
damage because all five perturbations preserve composition, and element-blind minimum-distance cutoffs detected almost none.

Law~1 replaces a fixed distance cutoff in angstroms with the radius-scaled reduced contact:
\begin{equation}
  \rhoc \equiv \min_{\mathrm{cation\text{--}anion\ contacts}}
  \frac{d}{r_{\mathrm{cat}}+r_{\mathrm{an}}},
  \label{eq:rho}
\end{equation}
where \(d\) is the contact distance and the denominator sums the cation and anion Shannon radii
\cite{shannon1976revised}. Cation and anion radii are assigned from composition-balanced formal charges and local coordination numbers. Because a fixed angstrom cutoff represents a larger fraction of the radius sum for small ions,
it penalizes small-ion contacts far more strictly than large-ion contacts. Law~1 enforces a lower bound \(\rhoc\geq\tau\), where the threshold \(\tau\) is calibrated to a chosen percentile of experimental crystals. Set~1 uses \(\tau=0.735\) (the first percentile
of the discovery distribution) and reached 99.2\% held-out satisfaction with 28.9\% damage
detection. If \rhoc{} captures short-range repulsion across chemistries, the energy cost of
compression relative to equilibrium should collapse onto \rhoc{}. We rigidly scaled twenty experimental compounds
along \rhoc{} and computed their plane-wave DFT energy above each compound's equilibrium minimum (Fig.~\ref{fig:anatomy}c). The DFT energy cost crossed
0.1\,eV per atom at median \rhoc{} = 0.927, and the spread of this crossing across compounds was 1.80 times narrower in
\rhoc{} than in \AA{}. Sets~1$'$--4 tighten the floor to \(\tau=0.804\) (the third discovery percentile). Both floors lie inside a compressed region that already
costs electronvolts per atom, with median costs of 5.36\,eV per atom at \(\tau=0.735\) and 2.08\,eV per atom at the stricter floor.

\begin{table*}[tp]
  \centering
  \caption{\textbf{The eight PRIS laws, their discovery-split percentiles, physical mechanisms and diagnostic meaning.} Here \fion{} is Pauling's composition-based ionic-character estimate, and \(\overline{\mathrm{CN}}_{\mathrm{an}}\) is the mean anion coordination number. The quantity \(\VM(i)\) is the site Madelung energy from an Ewald sum over formal charges \(z_i\) of magnitude \(v_i=|z_i|\), and BV denotes the bond valence in the bond-valence sum.}
  \label{tab:laws}
  \scriptsize
  \setlength{\tabcolsep}{2.5pt}
  \begin{tabular}{@{}l>{\raggedright\arraybackslash}p{0.30\textwidth}l>{\raggedright\arraybackslash}p{0.17\textwidth}>{\raggedright\arraybackslash}p{0.27\textwidth}@{}}
    \toprule
    Law & Mathematical statement & Discovery percentile & Mechanism & Violation points to \\
    \midrule
    Law 1 & \(\rhoc \geq \tau,\ \tau\in\{0.735,\,0.804\}\) & 1.0\% / 3.0\% & Short-range repulsion & Overlapping ions, wrong radius or charge \\
    Law 2 & \(\fion>0.50 \Longrightarrow \rhoc\leq1.05\) & 99.4\% (ionic) & Ionic contact & Expanded cell or rattling ion \\
    Law 3 & \(\overline{\mathrm{CN}}_{\mathrm{an}}\leq3.333 \Longrightarrow \overline{d/(r_{\mathrm{cat}}+r_{\mathrm{an}})}\leq1.081\) & 96.7\% (conditional) & Packing & Open, under-coordinated framework \\
    Law 4 & \(\operatorname{range}_i(\VM(i)/v_i)\leq31.45\,\mathrm{eV}\) & 99.5\% & Electrostatic balance & Incompatible site potentials \\
    Law 5 & \(\max_i \VM(i)\leq15.17\,\mathrm{eV}\) & 99.5\% & Electrostatic balance & Single strongly destabilized site \\
    Law 6 & \(\fion>0.55 \Longrightarrow \text{no like-charge bonds}\) & Discrete & Electrostatic balance & Cation--anion swap or mislabelled site \\
    Law 7 & \(n_{\mathrm{inequivalent\ sites}}/n_{\mathrm{sites}}\leq2/3\) & 91.6\% & Site simplicity & Artificial ordering or unsymmetrized cell \\
    Law 8 & \(\overline{|\mathrm{BV\ sum}-v_i|/v_i}\leq0.7143\) & 96.0\% & Bond-valence conservation & Mismatched element or oxidation state \\
    \bottomrule
  \end{tabular}
\end{table*}

Table~\ref{tab:laws} summarizes the eight laws, their discovery-split percentile thresholds, the underlying physical mechanisms and the structural failures diagnosed by each violation (with set memberships shown in Fig.~\ref{fig:loop}d). Laws~2, 3 and~6 are conditional and apply only within their stated ionicity or coordination regimes, so structures outside those regimes pass them by definition. When a new structure lacks the formal oxidation states or tabulated radii that a law requires, PRIS
returns an explicit no-verdict label rather than a default pass (details in Methods).

A lower bound on interatomic contact captures only short-range repulsion, leaving lattice expansion and coordinate-preserving cation--anion exchanges undetected. Because
distinct failure modes can leave the shortest contact unchanged, Set~2 incorporates rigid-ion packing and Madelung electrostatics,
and Set~3 extends the check to the charge-consistent bond network. Across the held-out split (Fig.~\ref{fig:rules}b), Set~3 achieved
91.7\% experimental satisfaction and 70.0\% pooled damage detection, exceeding 55.2\% in every perturbation class
and more than doubling the 28.9\% detection rate of Set~1.
Set~4 adds the bond-valence law (Law~8) and the distinct-site-fraction law (Law~7). Across the 5{,}297 held-out experimental and 3{,}612 damaged structures, Set~4 reached 81.8\% satisfaction and 91.1\%
overall detection, with at least 73.4\% in every class (Fig.~\ref{fig:rules}b and
Supplementary Figs.~S4 and~S5). Combining laws that target different failures enabled Set~4 to detect more damage without tightening one
cutoff.

Pauling's rules 2--5 and the five PRIS sets diverge sharply on the 5{,}297 held-out experimental structures (Fig.~\ref{fig:rules}c). Individual Pauling criteria were satisfied by 43.6\% (rule~2), 11.3\% (rule~3), 43.0\% (rule~4) and 81.6\% (rule~5) of held-out structures, and only 6.5\% satisfied rules 2--5 together \cite{pauling1929principles,george2020limited}. PRIS bridges the gap between overly restrictive Pauling rules and overly permissive distance cutoffs by grounding each inequality in an explicit physical mechanism while calibrating its threshold to the empirical tail of experimental crystals (Table~\ref{tab:laws}). Re-deriving each threshold at its defining percentile on the held-out split (Fig.~\ref{fig:rules}d) shifted the contact, packing and bond-valence thresholds by at most 1.5\%. For the two extreme-value electrostatic laws (Laws~4 and~5), which sit in the sparse 99.5th-percentile tail of Madelung energies (Supplementary Section~S18), re-derivation changed only 0.25--0.33\% of experimental verdicts and 5.0--7.8\% of damaged verdicts (Fig.~\ref{fig:rules}d, inset). Refitting Law~7 changed just 2.9\% of experimental and 3.2\% of damaged verdicts. Separating the physical mechanism from the population-calibrated threshold thus keeps each law interpretable and testable rather than reducing plausibility to an opaque fitted score. Even on held-out compositions absent from discovery (31.5\% of the held-out split), frozen Set~4 reached 82.7\% satisfaction and 91.9\% damage detection (Supplementary Figs.~S6--S11).

To evaluate how each layer of crystal chemistry improves screening on a common population, we benchmarked all baselines and Sets~1--4 on a separate set of 440 unrelaxed experimental structures and 2{,}024 damaged variants (details in Methods). Across the model sequence in Fig.~\ref{fig:rules}e, damage detection rose step by step as crystal chemistry was added. Composition-only charge neutrality detected 0\% of the damage, and the 0.5- and 0.7-\AA{} distance cutoffs used in generator pipelines \cite{xie2022crystal,betala2025lematgenbench} detected only 1.6--3.2\%. The radius-normalized floor (Set~1) and the two-sided ionic window (Set~1$'$) raised detection to 26.8\% and 35.8\%, and the rigid-ion lattice (Set~2) and the charge-consistent bond network (Set~3) lifted it to 58.8\% and 66.9\%. Full crystal chemistry in Set~4 detected 87.9\% of the damage at 83.0\% experimental satisfaction, 2.6-fold above a distance cutoff tightened to the same satisfaction. Resolving detection by perturbation class (Fig.~\ref{fig:rules}f) confirms that the laws generalize across failure modes and do not overfit a single distortion. Each mechanism detected both the perturbation classes used during law selection and classes withheld from selection
\cite{geirhos2020shortcut,torralba2011unbiased,lapuschkin2019unmasking}
(details in Methods and Supplementary
Figs.~S13 and~S14a,b).

\subsection{From screening to diagnosis: five complementary mechanisms}
\label{sec:mechanism}

\begin{figure*}[tp]
  \centering
  \includegraphics[width=\textwidth]{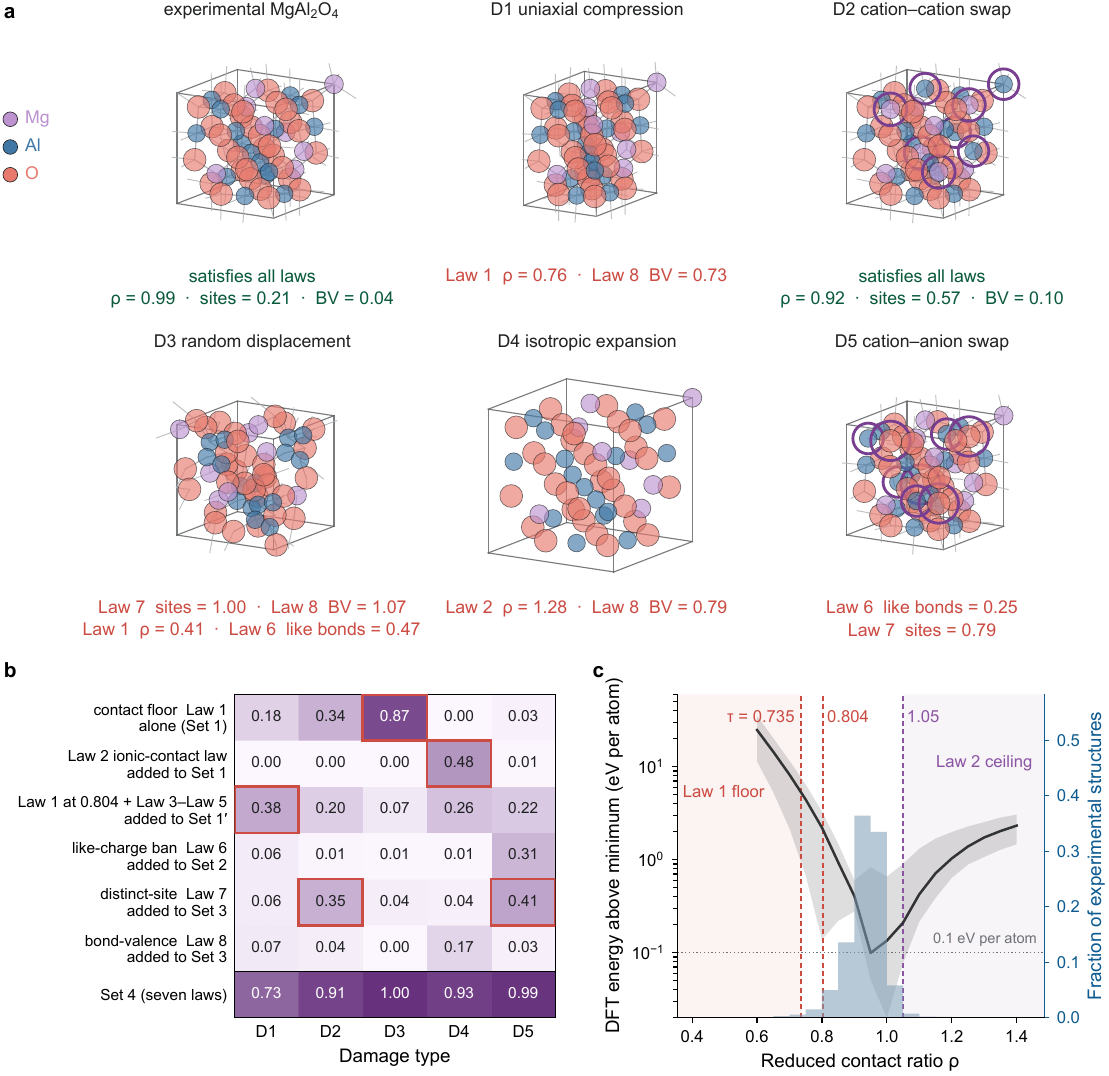}
  \caption[Physical basis of the laws]{\textbf{Physical basis of the PRIS plausibility
  laws.}
  \textbf{a},~Experimental MgAl$_2$O$_4$ and its five artificially damaged variants (uniaxial
  compression, cation--cation exchange, random displacement, isotropic expansion and
  cation--anion exchange). Relevant descriptors and law verdicts appear beneath each
  structure (green satisfied, red unsatisfied). Purple rings mark exchanged atoms.
  \textbf{b},~Held-out damage detection by perturbation class as laws are added,
  from Law~1 alone to Set~4. Red outlines mark the combinations referenced in the text.
  \textbf{c},~DFT energy above each compound's own minimum for twenty experimental
  compounds rigidly scaled along the reduced-contact coordinate \rhoc{}. Median and
  interquartile band (black, logarithmic left axis) are plotted together with the experimental
  distribution (blue, right axis). The two Law~1 floors and the conditional Law~2 ceiling
  are marked with their domains. The dotted line marks the 0.1\,eV per atom
  reference cost. Per-compound curves, the hard-potential check and the spread of the crossing in
  each coordinate are in Supplementary Fig.~S12.}
  \label{fig:anatomy}
\end{figure*}

Structural implausibility arises through five primary physical mechanisms, each tested by PRIS as an explicit inequality (Table~\ref{tab:laws}). Experimental spinel MgAl$_2$O$_4$ under five perturbations shows how the laws separate tolerable cation disorder, such as the Mg--Al exchange (D2) that approximates natural inverse-spinel order, from distinct failure modes (Fig.~\ref{fig:anatomy}a).
A 25\% uniaxial compression lowers \rhoc{} from 0.99 to 0.76 (violating Law~1), and a 30\% isotropic expansion raises \rhoc{}
to 1.28, above the 1.05 ceiling of Law~2. A
cation--anion exchange leaves \rhoc{} at 0.99 but creates eight like-charge bonds (violating Law~6). Each unsatisfied law
names both the measured quantity and its mechanism.

The first two mechanisms concern interatomic contact and coordination packing. At the shortest length scale, Law~1 encodes Born repulsion: Coulomb attraction varies as \(1/d\), whereas closed-shell repulsion rises
approximately exponentially under compression \cite{born1932gittertheorie} and places compressed contacts below \(\rhoc=0.804\) on a steep multi-eV wall
(Fig.~\ref{fig:anatomy}c). A lower bound cannot detect expansion, but in a
sufficiently ionic compound (where Pauling's composition-based ionicity \fion{} exceeds 0.50) electrostatic cohesion requires at least one cation--anion contact to approach the radius sum \cite{hawthorne2026crystal}. Law~2 therefore marks a
shortest contact beyond 1.05 times that sum as under-packed. Within this ionic domain, 96.25\% of
expanded discovery-split structures exceeded 1.05, compared with 0.62\% of
experimental ionic structures. Law~3 detects expanded low-coordination frameworks without Law~2's ionicity requirement by capping the mean
reduced cation--anion contact at 1.081 when the mean anion coordination number is at most 3.333.

Because geometry alone cannot detect an elemental swap that leaves atomic coordinates unchanged, Laws~4--6
evaluate electrostatic balance
\cite{ewald1921berechnung,hoppe1979effective}. Law~4 bounds the spread of charge-normalized site Madelung
energies (\(\VM(i)/v_i\leq31.45\)\,eV). Law~5 caps the highest site Madelung energy (\(\VM(i)\leq15.17\)\,eV) to flag a
single electrostatically destabilized site. For \fion{} above 0.55, Law~6 forbids first-shell like-charge
bonds and flags cation--anion exchanges that leave the distance matrix
unchanged.
Law~8 complements these long-range point-charge potentials by testing local valence matching through the bond-valence sum \cite{brown2009recent,brese1991bond}. Because bond valences decay exponentially with distance, expansion
lowers a site's valence sum, and compression or wrong-site substitution creates severe over- or under-bonding. Across discovery and held-out structures with tabulated parameters,
the structure-level mean relative bond-valence deviation had a median of 0.094 for experimental
crystals versus 0.705 for damaged structures. The experimental median corresponds to a global instability index of 0.215 valence units, matching the \(\sim0.2\) valence-unit convention in bond-valence analysis and showing that real crystals obey local valence matching within a \(\sim10\%\) tolerance.

The fifth mechanism counts distinct chemical environments: Law~7 caps crystallographically inequivalent sites at two-thirds of all unit-cell sites.
Displacing atoms or artificially splitting elements across equivalent Wyckoff orbits raises that fraction. By contrast,
91.6\% of experimental discovery structures keep it at or below \(2/3\). Even in an ablation that excluded all classical Pauling-derived descriptors (Wyckoff-site and space-group quantities, bond-valence sums and the measures of Pauling's rules 2, 4 and 5) from the discovery search, contact and charge-topology laws still retained about 80\% of the detection gain (Supplementary Section~S17). Stepping across the nested sets in Fig.~\ref{fig:anatomy}b shows how each added
mechanism fills the diagnostic gaps left by simpler crystal models. Together, they turn pass-or-fail screening into physical diagnosis by pinpointing whether a rejected structure
fails through ionic size, oxidation state, site occupancy, local bond valence or symmetry, judging plausibility before any geometry relaxation is run.

\subsection{Laws discovered without synthesis labels are highly linearly correlated with synthesizability}
\label{sec:certified}

\label{sec:pauling}\label{sec:database}\label{sec:scores}\label{sec:formulas}

\begin{figure*}[tp]
  \centering
  \includegraphics[width=\textwidth]{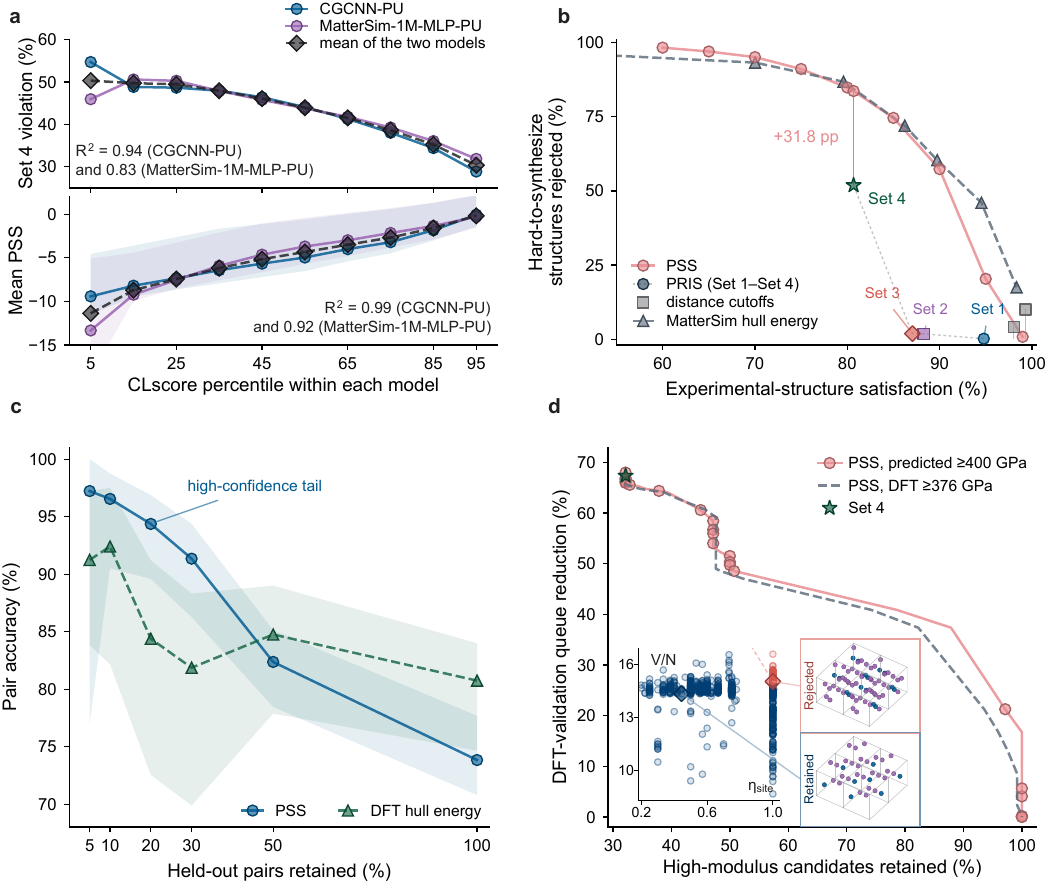}
  \caption{\textbf{Linear correlation with synthesizability, polymorph ranking and inverse-design screening.}
  \textbf{a},~Set~4 violation and mean PSS across within-model CLscore deciles for
  CGCNN-PU, MatterSim-1M-MLP-PU and their pointwise mean. Shading spans each model's
  interquartile PSS range. Each row gives the coefficient of determination (\(R^2\)) for each
  model's linear fit against the decile index.
  \textbf{b},~Satisfaction versus the fraction of hard-to-synthesize structures rejected.
  Curves compare Sets~1--4, the PSS threshold sweep, distance cutoffs (grey squares) and a
  sweep over a MatterSim-computed hull-energy threshold (slate-grey triangles). Connectors
  join Set~4 and PSS at matched satisfaction.
  \textbf{c},~Held-out same-composition pair accuracy for PSS and DFT
  \(E_{\mathrm{hull}}\) against the fraction of most-confident pairs retained. Shading
  gives confidence intervals.
  \textbf{d},~Across PSS thresholds, DFT-validation queue reduction versus retention of
  candidates that the universal machine-learned interatomic potential (UMA) predicts to
  have bulk modulus \(\geq400\)\,GPa. The Set~4 operating point is marked. The inset
  maps distinct-site fraction against atomic volume. Blue and red points denote retained
  and rejected candidates, respectively, and diamonds link a same-composition pair to the
  matching structure thumbnails. The dashed slate-grey curve uses the DFT-defined high-modulus
  subset at the mapped threshold of 376\,GPa among the 260 candidates evaluated from first
  principles (Supplementary Fig.~S20).}
  \label{fig:validation-synthesis}
\end{figure*}

If Laws~1--8 capture the structural regularities of real crystals, structures that violate them should also be harder to synthesize. Because the laws were derived without synthesis data, this prediction is an independent test. Because experimental databases record successful syntheses but rarely report failed attempts, standard binary classifiers cannot be trained directly. We therefore estimated synthesis likelihood using two contrasting positive--unlabelled (PU) machine-learning models. Each learns to output a continuous crystal-likeness score (CLscore) by contrasting known experimental crystals (positive examples) against a large unlabelled pool of candidate structures from the LeMat and ELEMENTA computational databases \cite{jang2020structure,song2025llm,ramlaoui2025lemat,elementa2025dataset}. The first model, CGCNN-PU, is a crystal graph convolutional network trained end-to-end on the synthesis task, and the second, MatterSim-1M-MLP-PU, trains a multilayer-perceptron classifier on structural features extracted from the pretrained MatterSim universal interatomic potential \cite{yang2024mattersim} (details in Methods). The 364{,}592 unique structures assigned the lowest CLscores by both models form the model-consensus hard-to-synthesize set. Evaluating Laws~1--8 against CLscore within each model independently (Fig.~\ref{fig:validation-synthesis}a, upper, and Supplementary Fig.~S18a--d) rules out shared-selection bias. Across within-model CLscore deciles, the mean Set~4 violation rate fell linearly from 50.3\% in the lowest decile to 30.4\% in the highest (linear-fit $R^2=0.94$ for CGCNN-PU and $0.83$ for MatterSim-1M-MLP-PU). Plausibility under laws discovered purely from crystal-chemical perturbations is thus highly linearly correlated with predicted synthesizability across the population, without any synthesis training labels.

To turn these discrete mechanisms into a continuous score for ranking polymorphs, our agents fitted the
PRIS-derived synthesis score (PSS) on same-composition pairs of experimentally recorded and computed-only
polymorphs in the development split (details in Methods). The fitted score is
\begin{equation}
\begin{aligned}
  \mathrm{PSS}(\mathbf{x})={}&-4.90\,\widetilde v_{\mathrm{atom}}
  -1.24\,\widetilde M_z-1.18\,\widetilde\Delta_{\mathrm{BV}} \\
  &-0.84\,\widetilde\eta_{\mathrm{site}}
  -0.22\,\widetilde k_{\max}+0.59\,\widetilde f_{\mathrm{iso}}, \\
  &\widetilde x=\frac{x^\dagger-\mu_x}{\sigma_x}.
\end{aligned}
\label{eq:pss}
\end{equation}
The six standardized terms (defined in Supplementary Section~S15) are atomic volume, site-Madelung environment, bond-valence
deviation, distinct-site fraction, maximum cation-polyhedron connectivity and the fraction
of isolated cation polyhedra. The negative weights on atomic volume, Madelung energy, bond-valence deviation and site complexity show that denser, electrostatically balanced, valence-matched and simpler polymorphs are more often synthesized. The negative weight on polyhedral connectivity and the positive weight on isolated polyhedra favour polymorphs whose cation polyhedra share fewer neighbours, in line with Pauling's third and fourth rules (half-sample stability in Supplementary
Figs.~S15 and~S16). Across CLscore deciles, mean PSS rose from $-11.37$ to $-0.18$, with linear-fit $R^2$ values of $0.99$ and $0.92$ in the two PU models (Fig.~\ref{fig:validation-synthesis}a, lower). On the consensus hard-to-synthesize set (Fig.~\ref{fig:validation-synthesis}b), Set~4 and PSS explainably rejected 51.9\% and 83.7\% of structures while retaining 80.7\% of experimental structures, with each penalty traceable to the physical terms in equation~\ref{eq:pss}. Both operate directly on unrelaxed coordinates, outperforming a MatterSim convex-hull energy threshold (\(E_{\mathrm{hull}}\), 72.0\% rejection) that requires geometry relaxation and a reference phase diagram (Supplementary Fig.~S17a,b).

Within a given composition, binary plausibility laws and continuous ranking serve complementary roles \cite{zeng2024selective}. Across held-out same-composition polymorph pairs, Set~4 frequently returned ties because multiple physically credible polymorphs satisfy all eight laws (Supplementary Fig.~S19a--c). Counting these ties as ranking failures makes Pauling's rules appear anti-predictive. An early agent analysis drew exactly that conclusion and proposed a symmetry mechanism to explain it, and within-composition refutation checks overturned both claims. Treated as non-decisions, the ties reveal a division of labour in which discrete laws define the boundary of the plausible region and continuous scores rank candidates inside it. Within that plausible region (Fig.~\ref{fig:validation-synthesis}c), PSS identified the experimentally recorded polymorph in 68.1\% of held-out pairs overall, compared with 75.0\% for DFT \(E_{\mathrm{hull}}\). On the top 20\% of pairs with the largest score differences between polymorphs under each criterion, PSS reached 94.4\% accuracy, surpassing DFT \(E_{\mathrm{hull}}\) at 84.4\% and allowing high-confidence polymorphs to be prioritized before costly DFT comparisons (details in Methods).

We then tested explainable, plausibility-first screening on 1{,}081 unique, unrelaxed candidates from MatterGen conditioned on a 400\,GPa bulk modulus, where the universal potential UMA \cite{wood2025uma} placed 140 candidates at or above 400\,GPa (Fig.~\ref{fig:validation-synthesis}d; details in Methods). For these predominantly intermetallic candidates, screening operates without formal charges through the site-simplicity law (Law~7) and the two charge-free PSS descriptors, distinct-site fraction and atomic volume, which weigh dense packing against site splitting (Fig.~\ref{fig:validation-synthesis}d, inset). At a conservative threshold retaining all 140 UMA-priority candidates, PSS removed 61 structures (red) that all passed the 0.7-\AA{} distance cutoff yet combined complete site splitting with open packing. In the matched Ir$_2$Os$_7$ pair, the rejected P1 candidate has 18 distinct sites among 18, \(V/N=15.017\)\,\AA$^3$ and PSS \(-0.638\), whereas the retained C2/m candidate has four distinct sites among nine, \(V/N=14.327\)\,\AA$^3$ and PSS \(1.245\). Plane-wave DFT bulk moduli for 260 candidates were a median 0.940 times the UMA predictions, mapping 400\,GPa to 376\,GPa on the DFT scale (with one candidate above 400\,GPa under DFT). Against that mapped threshold, a high-modulus candidate outranked a PSS-rejected candidate 96.6\% of the time, and PSS retained 99.2\% of the DFT high-modulus subset while tunable PRIS and PSS operating points cut the DFT validation queue by up to 67.3\% before hull-stability evaluation (Supplementary Fig.~S20). DFT relaxation also directly validated Law~7: as unconverged coordinates relaxed onto symmetry orbits, candidates satisfying Law~7 rose from 61 to 113 of 260, with none moving in the opposite direction and every new pass coming from PSS-retained candidates (Supplementary Fig.~S21). Where binary rules enforce fixed cutoffs on raw coordinates (with Set~4 cutting the queue by 67.3\% and charge-dependent Sets~1--3 by 0--0.4\%), continuous PSS balances packing density against unconverged site splitting to deliver an explainable queue reduction without losing target candidates. PRIS states why a structure is questionable, and PSS sets how strongly the queue is screened.

\subsection{Addressing controversy and identifying falsified structures in crystallography}
\label{sec:ordering-identity}

\label{sec:deploy}\label{sec:explain}

Beyond queue filtering, the same laws diagnose two structural controversies in existing databases: excess low-symmetry structures in generated databases \cite{merchant2023scaling,cheetham2024artificial,szymanski2023autonomous,leeman2024challenges,szymanski2026correction,yamazaki2026navigating} and wrong-element assignments at valid coordinates \cite{harrison2010falsified,iucr2012retraction}. Across Materials Project entries grouped by thermodynamic hull stability and dynamical stability (the absence or presence of imaginary phonon modes), 84--89\% of experimentally recorded crystals satisfied Law~7 in every stability class, compared with 43--66\% of unrecorded entries (Supplementary
Fig.~S22). A crystal can therefore pass distance cutoffs and sit near the convex hull while splitting chemically similar atoms across too many inequivalent sites.

\begin{figure*}[tp]
  \centering
  \includegraphics[width=0.97\textwidth]{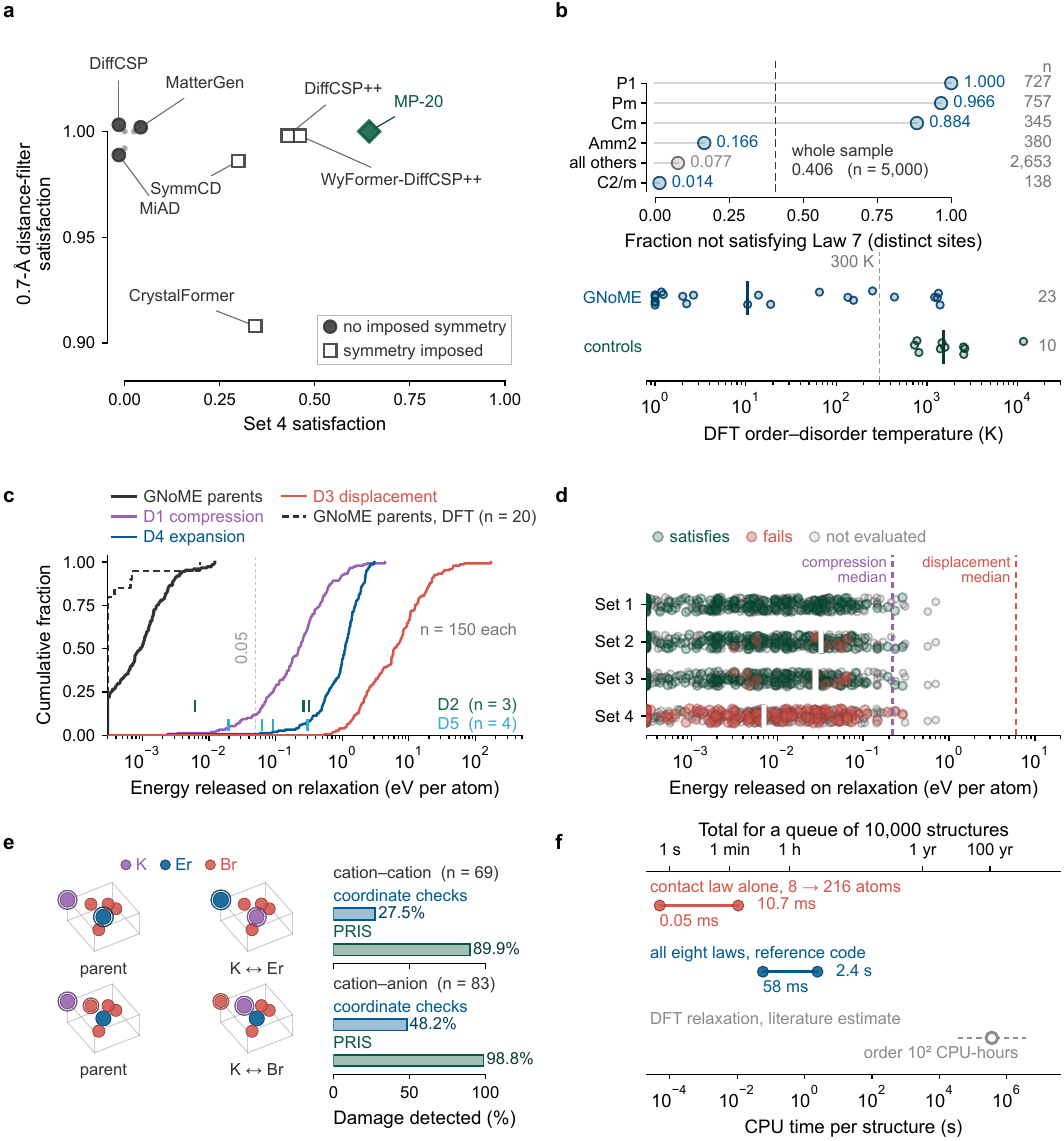}
  \caption{\textbf{Physicochemical screening of generated and database structures.}
  \textbf{a},~Minimum-distance-cutoff satisfaction versus Set~4 satisfaction among
  charge-assignable outputs from seven generators and from MP-20, the usual benchmark for
  these generators (green diamond). Filled circles mark generation without
  imposed symmetry, and open squares mark symmetry-constrained generation. Nearly
  coincident markers are offset and connected to their true coordinates.
  \textbf{b},~Upper, fraction of the GNoME sample failing Law~7 by source space group. The
  dashed line gives the whole-sample fraction. Lower, DFT order--disorder temperatures for
  23 label-merged GNoME entries and 10 experimental structures, obtained by relaxing every
  symmetry-distinct ordering of each entry. Vertical bars give class medians, and the dashed
  line marks 300\,K (Supplementary Fig.~S25).
  \textbf{c},~Cumulative distributions of the energy released on MatterSim relaxation for
  unmodified GNoME parents and their compressed, expanded and displaced counterparts.
  Sparse exchange classes appear as vertical marks. The dashed black curve gives the same
  quantity computed from first principles for twenty of these unmodified GNoME parents.
  First-principles relaxation energies for damaged cells were evaluated on
  experimental parents (Supplementary
  Fig.~S24).
  \textbf{d},~Relaxation-energy release for raw MatterGen outputs resolved by Sets~1--4
  verdict. White vertical segments with pale-grey outlines mark failure-group medians. The
  two coloured dashed lines mark the compression and displacement medians of panel c. The energy axis keeps panel c's logarithmic
  decade spacing over a shorter range.
  \textbf{e},~Fixed-coordinate cation--cation and cation--anion exchanges in one
  KErBr$_4$ parent. Bars compare four coordinate-based checkCIF/PLATON test families
  with the six PRIS laws on those exchanges.
  \textbf{f},~Measured CPU time per structure for Law~1 and all eight laws, with the
  literature DFT-relaxation estimate (dashed). The upper axis converts the same
  values to total time for the displayed queue.}
  \label{fig:deploy}
\end{figure*}

We first evaluated 500 unrelaxed structures from each of seven public crystal generators \cite{zeni2025generative,jiao2023crystal,okhotin2025miad,jiao2024space,cao2025space,levy2025symmcd,kazeev2025wyckoff} against MP-20, the standard Materials Project benchmark of crystals with up to 20 atoms per unit cell (Fig.~\ref{fig:deploy}a; details in Methods). The 0.7-\AA{} distance cutoff passed nearly all outputs indiscriminately. By contrast, Set~4 satisfaction among charge-assignable outputs was 0--2.5\% for generators that place atoms freely, 29.9--46.1\% for generators that place atoms on Wyckoff positions, and 64.3\% for MP-20. Law~7 accounted for much of this gap even after geometry relaxation and symmetry re-detection, showing that Wyckoff-constrained generation is crucial for producing crystallographically plausible structures.

Because Law~7 needs no charge assignment, we audited 5{,}000 uniformly sampled GNoME structures (details in Methods). Overall, 40.64\% failed Law~7 (Fig.~\ref{fig:deploy}b, upper), including all \(P1\) structures and nearly all \(Pm\) and \(Cm\) structures (charge-law coverage in Supplementary Fig.~S23). This low-symmetry excess could reflect unrelaxed mechanical strain or artificial chemical ordering across sites that should be equivalent. To test the first explanation, we compared cumulative relaxation-energy distributions under MatterSim \cite{yang2024mattersim} and DFT (Fig.~\ref{fig:deploy}c). For 150 unmodified GNoME parents, the median MatterSim relaxation release was below 0.001\,eV per atom (0.0001\,eV per atom under DFT for twenty parents; Spearman rank correlation 0.953 across 200 DFT-checked cells). Compressed and displaced (0.3--0.8\,\AA{}) GNoME counterparts instead released medians of 0.22 and 6.05\,eV per atom, respectively. Both damage medians align with the 3.15\,eV per atom DFT median (3.33\,eV per atom in MatterSim) for displaced experimental parents, where random 0.3--0.8\,\AA{} shifts push ion pairs into short-range Born repulsion (Supplementary Fig.~S24). Across 500 raw MatterGen outputs grouped by PRIS verdict in Fig.~\ref{fig:deploy}d, structures failing Set~4 released a median of only 0.007\,eV per atom, close to the 0.006\,eV per atom of the no-verdict group. A small relaxation energy therefore confirms only local mechanical equilibrium while leaving electrostatic, bond-valence and site-complexity errors undetected.

With unrelaxed strain ruled out, we next tested whether artificial cation ordering created GNoME's extra sites by merging chemically similar elements (such as rare-earth pairs or Zr--Hf) under one shared label at fixed coordinates. Among 150 sampled Law~7-failing GNoME entries, 113 contained a mergeable pair. At a 0.1-\AA{} symmetry tolerance, relabelling brought 78\% of the 113 entries within the Law~7 threshold and increased space-group symmetry in 79\%, a change seen in none of 27 mergeable experimental structures. We then relaxed every symmetry-distinct site permutation with DFT for 23 such GNoME entries and 10 experimental structures (Fig.~\ref{fig:deploy}b, lower, and Supplementary Fig.~S25). Scrambling a GNoME ordering cost a median energy of only 0.0001\,eV per atom, compared with 0.036\,eV per atom for experimental crystals. The orderings were energetically interchangeable below 300\,K in 18 of the 23 GNoME entries and in none of the experimental structures. These near-zero ordering energies provide direct quantitative support for the critique that generated databases substitute nearly indistinguishable elements \cite{smit2026dataonly}. Such energies are consistent with a thermodynamic reading of Law~7 and Pauling's fifth rule: a site distinction that carries negligible ordering energy cannot persist against thermal disorder at room temperature.

In the converse failure mode, atomic coordinates remain valid while the wrong chemical element is assigned to a crystallographic site. Harrison and colleagues documented at least 70 falsified crystal reports in which genuine diffraction intensities were paired with altered element symbols, originally caught retrospectively through Hirshfeld rigid-bond alerts and structure-factor checks \cite{harrison2010falsified}. We held the lattice and all fractional coordinates fixed while swapping chemical species in experimental parents (Fig.~\ref{fig:deploy}e and Supplementary Section~S17). Four checkCIF/PLATON test families that can be computed from coordinates alone \cite{spek2009structure} detected 19 of 69 cation--cation exchanges (27.5\%) and 40 of 83 cation--anion exchanges (48.2\%), whereas PRIS detected 62 of 69 (89.9\%) and 82 of 83 (98.8\%). In a recovered archive of retracted depositions \cite{iucr2012retraction}, four evaluable entries share one M(C$_2$H$_2$N$_3$Cl) framework with M labelled Cu, Ni, Mn or Fe. All four entries gave \(\rhoc=0.53\) and mean bond-valence deviations of 0.77--0.78, violating Laws~1 and~8 because the substituted ions are too large for the fixed site distances and fail local valence matching.

These diagnostic checks require a negligible fraction of the computation needed for first-principles validation. Benchmarking CPU time per structure (Fig.~\ref{fig:deploy}f) shows that evaluating 10{,}000 unit cells of 6--20 atoms takes under one second for Law~1 alone and about ten minutes for all eight PRIS laws. Full PRIS evaluation thus costs less than one millionth of the estimated 100\,CPU-hours per DFT relaxation (one million CPU-hours for a 10{,}000-structure queue) \cite{jain2013commentary}, allowing generators, structure-prediction workflows and autonomous laboratories to run an explainable diagnosis before expensive validation.

\section{Discussion}
\label{sec:ledger}

Experimental inorganic crystals obey five quantitative constraints on short-range repulsion, ionic contact and packing, point-charge electrostatics, local bond valence and crystallographic site simplicity (Table~\ref{tab:laws}). Each constraint bounds an explicit structural descriptor at a calibrated percentile of experimental crystals. Where the ionic model requires it, the bound is gated by an ionicity condition, and the short-range contact bounds are anchored to a steep DFT repulsive energy wall. On the same held-out structures, Pauling's rules 2--5 together were satisfied by 6.5\% of experimental structures, whereas the PRIS law sets reached 82--99\% \cite{pauling1929principles,george2020limited}. PRIS thus provides the quantitative empirical rules sought beyond Pauling's \cite{george2020limited} while complementing ongoing quantum-mechanical and bond-topological reformulations of crystal chemistry \cite{gibbs2022pauling,hawthorne2026crystal}.

Our discovery, benchmark and external tests establish structural plausibility as
a distinct first layer of crystal assessment that precedes thermodynamic stability, dynamical
stability and synthesizability. Downstream stability and synthesis metrics do not track one another. On published phonon data
\cite{petretto2018highthroughput,zhu2024highthroughput}, 35.6\% of on-hull structures had
imaginary modes, 41.7\% of experimentally recorded entries were metastable, and 4{,}271
unrecorded entries were dynamically stable and within 50\,meV per atom of the hull (details in Methods).
More fundamentally, none of these downstream calculations tests whether the proposed crystal structure is physicochemically self-consistent in the first place
\cite{sun2016thermodynamic,aykol2018thermodynamic,antoniuk2023predicting}. Because plausibility, stability and synthesizability address distinct physical questions that cannot be collapsed into a single
composite score, PRIS provides an explicit, falsifiable diagnosis before downstream evaluation.
Moreover, the five plausibility mechanisms reach well beyond static geometry. Plausibility under laws discovered without synthesis labels is highly linearly correlated with predicted synthesizability at the population level across two contrasting PU models, and the fitted PSS weights confirm that high density, balanced electrostatics, satisfied bond valences and simple site symmetry favour synthesis.

PRIS acts at three levels of decision: the validation queue, the database and the
individual site. At the queue level, population-derived conditional laws bridge the
restrictive Pauling rules and permissive distance cutoffs. A binary pass-or-fail screen appropriately
treats competing viable polymorphs as equally plausible, leaving continuous scores such as PSS or DFT energy to rank them. In inverse design, PSS weighs dense packing against site splitting to explainably cut the DFT queue by up to 67.3\% while retaining 99.2\% of the DFT-verified high-modulus candidates. At the database
level, Law~7, similar-element relabelling and near-zero DFT permutation energies trace GNoME's low-symmetry excess to artificial site
splitting that thermal energy would disorder at room temperature. At the individual-site level, charge- and bond-valence-sensitive laws expose elemental
misassignments and falsified crystal reports at fixed coordinates that purely geometric checks cannot detect, while assigning a separate no-verdict label when formal charges or radii are unavailable.

More broadly, our results demonstrate how active refutation can turn autonomous agents from
prolific hypothesis generators into self-correcting scientific discoverers
\cite{boiko2023autonomous,szymanski2023autonomous,mbran2024augmenting}. Operating within
human-defined goals and data partitions across 572
investigations and 2{,}037{,}606 candidate evaluations, our agents subjected every promising
rule to adversarial counterexamples and physical controls, rejecting eleven seemingly successful
conclusions to distil eight compact laws. Preserving the complete failure record
links each surviving law to the refutation tests it withstood, fulfilling calls to report
failed attempts as a condition for progress
\cite{smit2026dataonly}. Grounded in explicit physical mechanisms and an auditable
falsification record, PRIS turns crystal screening from a pass-or-fail filter into a diagnosis of why a
candidate structure is implausible. Coupled with systematic refutation, autonomous agents can distil
compact physicochemical laws from crystallographic data to guide both computation and synthesis.

\section{Methods}

Supplementary Section~S21 gives these Methods in full.

\subsection{Autonomous-agent workflow, model and human oversight}

The autonomous programme ran from 27 July to 14 August 2026. Our agents reasoned entirely
by calling Codex through our own \texttt{codex-api} client (\url{https://github.com/szl666/codex-api}).
They proposed hypotheses, implemented analysis scripts and ran diagnostic checks on failed
claims. Human researchers defined the initial objective, redirected the study scope and controlled
access to held-out partitions without supplying a detailed scientific hypothesis. They
verified all final outputs and retain responsibility for the reported conclusions.
Supplementary Sections~S1, S6, S9,
S14 and~S19 report the agent records, eleven refuted
conclusions and reproduction commands.

\subsection{First-principles verification}

Four quantities first learned from data were re-derived with plane-wave DFT: the contact
thresholds of Laws~1 and~2, the ordering behind GNoME's low-symmetry excess, the machine-learned
damage-severity scale and the design-queue property target. Plane-wave calculations were performed in VASP 6.3.0
using PBE PAW potentials across 1{,}917 tasks to obtain rigid-scaling energy curves
along \rhoc{}, site-permutation ordering energies for element-merged structures, matched-cell relaxation
energies and Birch--Murnaghan bulk moduli. Supplementary Section~S20 gives
the protocol, and relaxed cells for the 260 design candidates are provided as
Supplementary Data.

\subsection{Data sets and study design}

Written protocols fixed the allowed structural quantities, data splits and success
criteria before evaluation. A seeded hash of each identifier assigned eligible
charge-balanced ionic structures among 99{,}162 ICSD and COD experimental structures
\cite{zagorac2019recent,grazulis2009crystallography} to discovery, held-out or reserve
partitions. Thresholds were
fitted on 12{,}632 experimental and 8{,}590 damaged discovery structures, then assessed
without refitting on 5{,}297 experimental and 3{,}612 damaged held-out structures (Supplementary
Sections~S2, S8 and~S19).

\subsection{Structural descriptors and law-set evaluation}

Formal oxidation states were assigned from composition by integer charge balancing
\cite{jablonka2021using}, which defines cations and anions by formal-charge sign. Primary
neighbours from CrystalNN \cite{zimmermann2020local} in pymatgen \cite{ong2013python}
supplied coordination numbers for radius and bond-valence lookups. When an (element,
oxidation state, coordination number) entry was absent from Shannon's table
\cite{shannon1976revised}, the lookup stepped through \(\mathrm{CN}-1\), \(\mathrm{CN}+1\),
sixfold coordination, the element's mean ionic radius, and \(1.0\)\,\AA{} for a cation or
\(1.4\)\,\AA{} for an anion. Law~1 used \(\tau=0.735\) in Set~1 and \(\tau=0.804\) in
Sets~1$'$--4, and Law~7 used spglib \cite{togo2024spglib} with \texttt{symprec}=0.01.
Within the benchmark, an unavailable measurement counted as satisfying the law. In
deployment, a structure without an assignable formal charge or a required feature receives a separate no-verdict label
(Supplementary Sections~S3, S17 and~S18).

\subsection{Artificially damaged structures and law selection}

The artificially damaged structures were generated from experimental parents through five
composition-preserving perturbations: 15--30\% uniaxial compression, swapping cations of
different formal charges, random Gaussian atomic displacement with a standard deviation of
0.3--0.8\,\AA{}, 20--40\% isotropic expansion and swapping cations with anions. We
selected laws to maximize damage detection subject to an experimental-structure
satisfaction floor, then evaluated fixed Set~4 once on held-out data and compared it with a
provably optimal decision tree of depth three. Leave-one-damage-class-out evaluation repeated
the selection with each class withheld. The distance-cutoff
benchmark used 440 experimental structures and 2{,}024 damaged variants (Supplementary
Sections~S4, S5, S7, S10,
S11, S12, S17 and~S18).

\subsection{External evaluation, synthesizability screening and statistics}

With PRIS thresholds fixed, we evaluated 500 outputs from each of seven crystal generators and sampled 5{,}000 of the 554{,}054 released GNoME structures. We also relaxed 500 raw MatterGen outputs and compared
26{,}600 Materials Project structures. Same-composition ranking covered 18{,}920 pairs, with composition-cluster bootstrap intervals. PSS was fitted
on development compositions as an antisymmetric, zero-intercept logistic score of the six
descriptors in equation~\ref{eq:pss}. A CGCNN-PU model and MLP PU heads on frozen MatterSim
representations \cite{jang2020structure,song2025llm,yang2024mattersim} defined the
hard-to-synthesize set by consensus. For inverse design, 13 MatterGen runs conditioned on a
400\,GPa bulk modulus yielded 1{,}081 unique candidates, and UMA predictions defined the
high-modulus subset (Supplementary Sections~S13, S15,
S16, S17 and~S19).

\section*{Supplementary data}
Supplementary Information and the DFT-relaxed structures are available as ancillary files with this arXiv submission.

\section*{Data availability}
Entries from the Inorganic Crystal Structure Database are not redistributable under
the terms of the FIZ Karlsruhe licence and the European Union \emph{sui generis}
database right, and ELEMENTA is distributed under CC-BY-NC-4.0. The public benchmark
released with this work is therefore built exclusively on the Crystallography Open
Database (CC0). Derived scalar features, split assignments and all numerical results
underlying the figures are provided as source data. The 260 design candidates
relaxed with density functional theory in this work are provided as Supplementary
Data: one CIF per candidate, with an index giving its composition, its space
group and distinct-site fraction both as generated and after relaxation, its
synthesizability score, and its computed bulk modulus.

\section*{Code availability}
All analysis code, the pre-registration document and the figure-generation scripts
are available at \url{https://github.com/AI4QC/PRIS}.

\section*{Acknowledgements}
We thank HKUST Fok Ying Tung Research Institute and National Supercomputing Center in Guangzhou Nansha
Sub-center for computational resources.

\section*{Funding}
This work was supported by the Start-up Fund of the Hong Kong University of Science and Technology.

\section*{Author contributions}
Z.S. and L.C. conceived the study. Z.S. built the agent workflow, analysed the results,
performed the density functional theory calculations and wrote the manuscript.
H.L. provided insights and revised the manuscript. L.C.
supervised the study and revised the manuscript.

\par\medskip\noindent\textit{Conflict of interest statement.} None declared.

\bibliographystyle{nsr}
\bibliography{refs}

\end{document}